\documentclass[prb, twocolumn, superscriptaddress, nofootinbib]{revtex4-2}
\usepackage{latexsym}
\usepackage{amssymb}
\usepackage{amsmath}
\usepackage{bm}
\usepackage[dvipdfmx]{graphicx}
\usepackage{color}

\begin{document}
\title{Diffusion model for analyzing quantum fingerprints in conductance fluctuation}
\author{Naoto Yokoi}
    \affiliation{Department of Applied Physics, The University of Tokyo, Tokyo 113-8656, Japan}
    \affiliation{Institute for AI and Beyond, The University of Tokyo, Tokyo 113-8656, Japan}
\author{Yuki Tanaka}
    \affiliation{Department of Applied Physics, The University of Tokyo, Tokyo 113-8656, Japan}
\author{Yukito Nonaka}
    \affiliation{Faculty of Science and Engineering,  Aoyama Gakuin University, Kanagawa 252-5258, Japan}
\author{Shunsuke Daimon}
    \affiliation{Quantum Materials and Applications Research Center, National Institutes for Quantum Science and Technology, Tokyo 152-8550, Japan}
\author{Junji Haruyama}
    \affiliation{Faculty of Science and Engineering,  Aoyama Gakuin University, Kanagawa 252-5258, Japan}
    \affiliation{Institute for Industrial Sciences, The University of Tokyo, Tokyo 153-8505, Japan}
\author{Eiji Saitoh}
    \affiliation{Department of Applied Physics, The University of Tokyo, Tokyo 113-8656, Japan}
    \affiliation{Institute for AI and Beyond, The University of Tokyo, Tokyo 113-8656, Japan}
    \affiliation{Advanced Institute for Materials Research, Tohoku University, Sendai 980-8577, Japan}
    \affiliation{RIKEN Center for Emergent Matter Science (CEMS), Wako 351-0198, Japan}
\begin{abstract}
A conditional diffusion model has been developed to analyze intricate conductance fluctuations 
called universal conductance fluctuations or quantum fingerprints appearing in quantum transport phenomena.  The model reconstructs impurity arrangements 
and quantum interference patterns in nanometals by using magnetoconductance data,  
providing a novel approach to analyze complex data based on machine learning.      
In addition, we visualize the attention weights in the model, which efficiently 
extract information on the non-local correlation of the electron wave functions,  
and the score functions, which represent the force fields in the wave-function space.   
\end{abstract} 

\maketitle

\renewcommand{\thefootnote}{\arabic{footnote}}
\setcounter{footnote}{0}

\section{Introduction} 
In the traditional approach to physics, only signals that could be understood by humans 
were targeted, while complex signals outside this range were labeled as noise (or fluctuation)  
and, with the exception of statistical analysis, were rarely considered subjects of scientific study. 
However, such complex signals are likely to contain rich information 
about the physical system, and this information has been essentially discarded. 
Many noise-like signals are generated from complex quantum systems, and 
understanding these signals and extracting useful information from them 
is a crucial step in advancing quantum science and materials science. 

In the field of data science, there have been remarkable advances   
in the analysis of complex data using machine learning (ML) and 
artificial intelligence (AI).  
In particular, the innovative development of generative models, such as large language models 
and diffusion models, suggests the potential to transform traditional methods of 
human cognition and understanding. 
(For recent developments in ML and generative models, see \cite{pml1Book, pml2Book, bishop2023deep}.)
In light of these developments, it is anticipated that the use of AI will enable 
new approaches to the analysis and understanding of complex signals in quantum systems. 
As an example of AI-based study on complex signals in quantum systems, 
we construct a diffusion model \cite{sohl2015deep, ho2020denoising} to decode 
the complex conductance fluctuations observed in quantum transport \cite{datta1997electronic}. 

Quantum transport is a phenomenon observed in nanoscale devices, 
where the quantum mechanical properties of electrons become pronounced, 
leading to conduction that preserves quantum coherence. In particular, the interference 
between wave functions scattered by impurities or reflected from the edges of the device 
causes the magnetoconductance to exhibit highly complex fluctuation patterns 
under the applied magnetic field. These complex fluctuation patterns are governed 
by the Schr\"{o}dinger equation, distinctly different from classical random fluctuations 
such as thermal noise. These patterns retain microscopic information, 
such as the positions of impurities, and are commonly referred to as ``quantum fingerprints.'' 
As an initial step toward understanding the nature of quantum fingerprints, 
a prototype generative model based on a variational autoencoder (VAE) has been developed  \cite{daimon2022deciphering}. 
This model successfully deciphered quantum fingerprints 
associated with restricted configurations of two impurities in devices with high accuracy. 
However, its performance is limited when applied to larger datasets of quantum fingerprints 
arising from general impurity configurations.  

In this letter, we aim to bridge quantum fingerprints and the microscopic structures of devices   
by constructing a diffusion model capable of deciphering more general quantum fingerprints.    
Using this model, we fully reconstruct impurity arrangements 
and their resulting quantum interference patterns from quantum fingerprint data 
associated with general configurations of two impurities. Quantum fingerprints are incorporated 
into the diffusion model as conditional inputs related to the interference patterns. 
Unlike the VAE-based approach, which requires a two-stage learning process, 
the learning in the diffusion model is unified into a single stage.

\section{Diffusion models as score-based generative models}  
We consider the data space $\bm{{\cal X}}$, which is a $D$-dimensional 
vector space $\bm{R}^{D}$. The data is denoted as a vector $\bm{x} \in \bm{{\cal X}}$, 
whose components are given by $x_{i}$ ($i=1, \ldots, D$). 
Generative models infer the probability distribution $P(\bm{x})$ 
on the data space $\bm{{\cal X}}$ through machine learning and generate 
unknown (but similar) data by sampling from the inferred 
probability distribution. Among them, score-based generative models
make inferences based on score functions, which are the gradients of
the probability distribution, $\bm{s}(\bm{x}) = \nabla_{\bm{x}}\log P(\bm{x})$, 
instead of using $P(\bm{x})$ itself \cite{hyvarinen2005estimation}. Note that by utilizing 
the Boltzmann form of the probability distribution, $P(\bm{x}) = e^{-\beta E(\bm{x})}/Z$, 
the score functions can be identified as the force field, 
$\bm{F}(\bm{x}) = - \beta\,\nabla_{\bm{x}} E(\bm{x})$, 
where $Z$ is the partition function (or normalization constant) 
and $\beta=1/k_{B} T$ is the inverse temperature.  

Diffusion models \cite{sohl2015deep, ho2020denoising} are a type of score-based generative 
model that learns the score functions through the denoising processes.   
First, we consider perturbed data by adding noise, 
$\tilde{\bm{x}} = \bm{x} + \bm{\epsilon}$, where $\bm{\epsilon}$ follows 
a $D$-dimensional normal distribution, $P(\bm{\epsilon}) \propto {\cal N}(\bm{0}, \sigma^{2}\bm{I})$, 
with zero mean and variance $\sigma^{2}$ ($\bm{I}$ is the identity matrix).  
Next, we introduce a deep neural network model $\bm{s}_{\bm{\theta}}(\tilde{\bm{x}})$,  
which approximates the score function with model parameters $\bm{\theta}$. 
Following the denoising score matching strategy \cite{vincent2011connection}, the model 
$\bm{s}_{\bm{\theta}}(\tilde{\bm{x}})$ is trained through machine learning 
based on a loss function, 
$L(\theta) = \mathbb{E}_{P(\bm{\epsilon}),\, P(\bm{x})} \left[ \left| 
\bm{\epsilon}/\sigma^{2} + \bm{s}_{\bm{\theta}}(\bm{x} + \bm{\epsilon}) \right|^{2} \right]$, 
where $\mathbb{E}_{P(\bm{z})}$ represents the expectation value with respect to 
the probability distribution $P(\bm{z})$. 
This shows that denoising, or inferring the noise, yields the desired score function.  
The diffusion model generates data through sampling based on the Langevin 
Monte Carlo method using the score function obtained from the denoising process  \cite{welling2011bayesian}. 

In order to generate specified data (such as images) given a condition (such as sentences), 
one should consider conditional diffusion models \cite{dhariwal2021diffusion, ho2022classifier} that provide the score function 
of the conditional probability distribution: $\nabla_{\bm{x}} \log P(\bm{x}\,|\, y)$,  
where $y$ represents the appropriate condition. 
In the actual learning process, deformed score functions of 
the conditional probability distribution are taken into account: 
$\nabla_{\bm{x}}  \log P_{\gamma}(\bm{x}\,|\, y) = \gamma \nabla_{\bm{x}} \log P(\bm{x}, y) 
+ (1-\gamma) \nabla_{\bm{x}} \log P(\bm{x})$, 
where $P(\bm{x}, y)$ is the joint probability distribution for the data $\bm{x}$ and the condition $y$, 
and $\gamma$ is the deformation parameter. The learning of this deformed 
score function can be implemented simply by concatenating the data 
with the condition during the learning process.

\section{Learning dataset}  
\begin{figure}
\includegraphics[width=8.6cm]{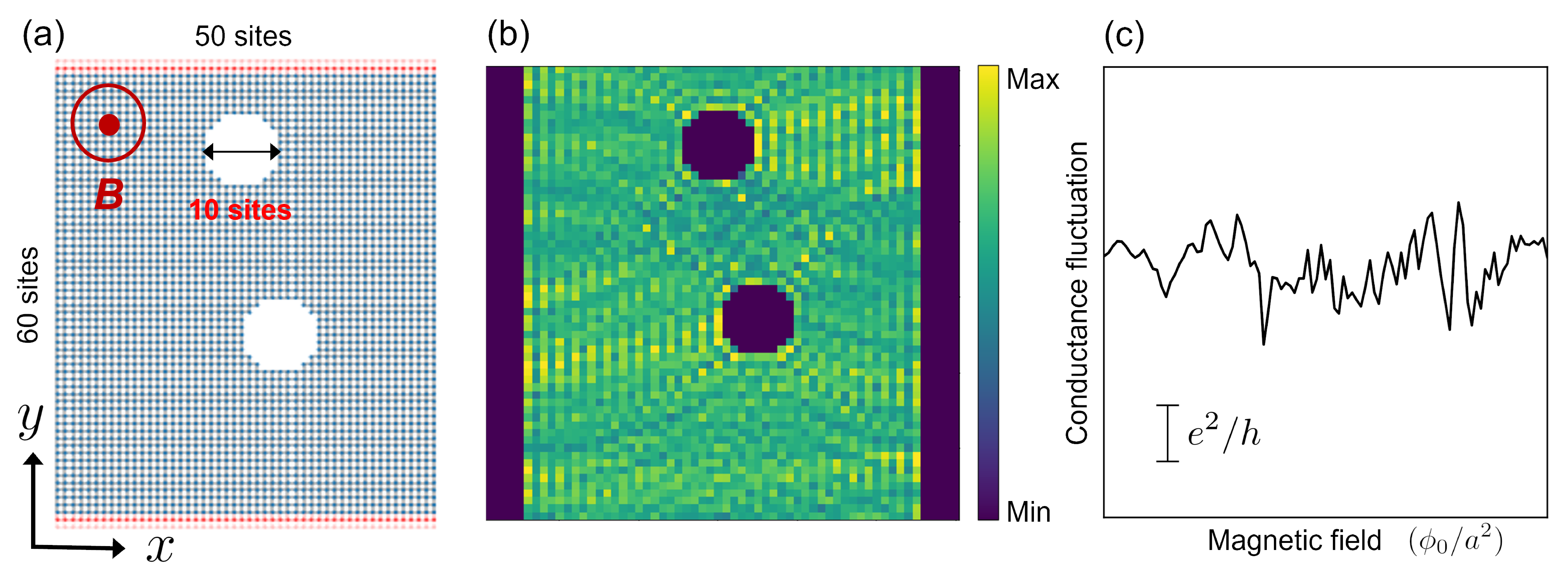}
\caption{Learning dataset: (a) lattice setup for numerical calculation, (b) LDOS determined by wave functions,  
(c) quantum fingerprint (magnetoconductance).}
\end{figure}
For the preparation of the learning dataset we performed numerical calculations 
of the electric conductance under a perpendicular magnetic field (magnetoconductance) 
for two-dimensional metal nanowires. Our calculation models of the nanowires 
consist of a square lattice with dimensions of $60 \times 50$ sites, and for simplicity, defects are 
introduced as two circular antidots, each with a radius of $5$ sites. 
This model, with a rectangular shape and infinitely long leads in the longitudinal $y$-direction,  
exhibits geometric symmetries: translation in the $y$-direction and inversion about 
the $x$-axis and $y$-axis. These symmetries result in multiple distinct configurations of the antidots 
that produce the same magnetoconductance. By carefully investigating these symmetries, 
we selected 16,170 antidot configurations corresponding to unique conductance data and 
numerically calculated the magnetoconductance for the selected configurations 
under five different small random potentials.  
The wave functions in the nanowires with the antidot configurations are simultaneously calculated, 
and the local density of states (LDOS) of electrons, given by the absolute square of 
the wave functions, is obtained 
under the condition of zero magnetic flux density. 
See FIG. 1 for an example.\footnote{For simplicity, we add zero padding 
with a width of  $5$ pixels to the left and right edges of each image of the interference patterns, 
resulting in a final size $60 \times 60$ pixels.}  

The numerical calculations are performed using a software package for  
calculating quantum transport properties called Kwant \cite{groth2014kwant}, 
which is based on two-dimensional tight-binding models under perpendicular magnetic fields \cite{ando1991quantum, datta1997electronic}.     
In total, we obtain 80,850 pairs of magnetoconductance and LDOS data, 
which are prepared as a learning dataset. This dataset is divided into 56,595 training data and
24,255 validation data. 

It is worth noting that the learning dataset 
in this study is substantially larger than that used in the previous VAE-based study \cite{daimon2022deciphering}, which focused exclusively on restricted configurations 
with a fixed position of the upper dot, and the accuracy of interference reconstruction 
using the VAE-based model drops to around $60$ percent for this dataset.

\section{Learning quantum transport by diffusion model}  
The network architecture for the denoising diffusion model used to decode the 
quantum fingerprints is primarily based on the U-Net architecture, a convolutional 
neural network (CNN) designed for image segmentation \cite{ronneberger2015u}. 
The U-Net consists of an encoder network with two-stage downsampling and 
a decoder network with two-stage upsampling,  
both constructed from multi-layer convolutional blocks.  
In addition, we introduce transformer blocks based on the self-attention mechanism 
to improve learning performance \cite{vaswani2017attention}.  
In order to generate interference patterns of LDOS 
from the corresponding magnetoconductance, 
we construct a conditional diffusion model, conditioned on the magnetoconductance data. 
To implement this conditioning, the one-dimensional magnetoconductance data are reshaped 
to match the size of the interference images using an additional transformer network. 
The reshaped data are then concatenated with the input vectors 
in the transformer block at each layer of the U-Net. 
The loss function is defined as the mean squared error, 
and the Sharpness-Aware Minimization (SAM) optimizer is employed \cite{foret2020sharpness}.    
The full network architecture is shown in FIG. 2. 
\begin{widetext}
\begin{center}
\begin{figure}[ht]
\includegraphics[width=16.2cm, height=5.8cm]{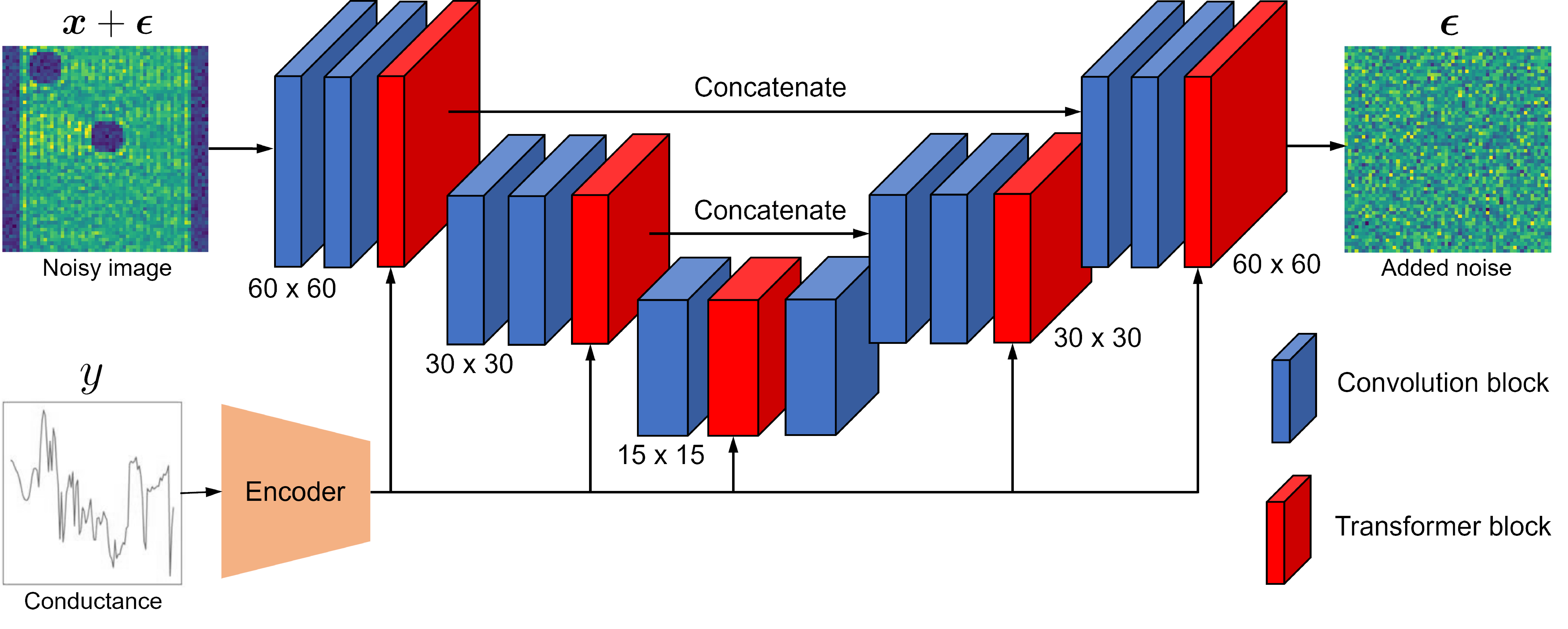} 
\caption{Schematic picture of denoising network of diffusion model.} 
\end{figure}
\end{center}
\end{widetext}
We set 1,000 diffusion steps for the training and generation processes in the diffusion model.   
During the training, the number of steps is randomly sampled from a uniform 
distribution, and the model parameters are optimized at each sampled step.  

Using the diffusion model described above, we successfully generated 
the correct quantum interference patterns of the wave function (or LDOS) 
and accurately reproduced the positions of antidots 
based on the magnetoconductance data. 
FIG. 3 illustrates the process in which the interference pattern is generated 
using the Langevin Monte Carlo method, based on the score functions 
obtained from the learning process. 
This generation is achieved by conditioning the magnetoconductance data, 
and it starts from almost random noise at the initial stage (step 1). 
We confirmed that our diffusion model can reproduce the interference patterns and 
the positions of the antidots with high accuracy (over $95$ percent) for other 
validation data as well.  
\begin{figure}[ht]
\includegraphics[width=8.6cm]{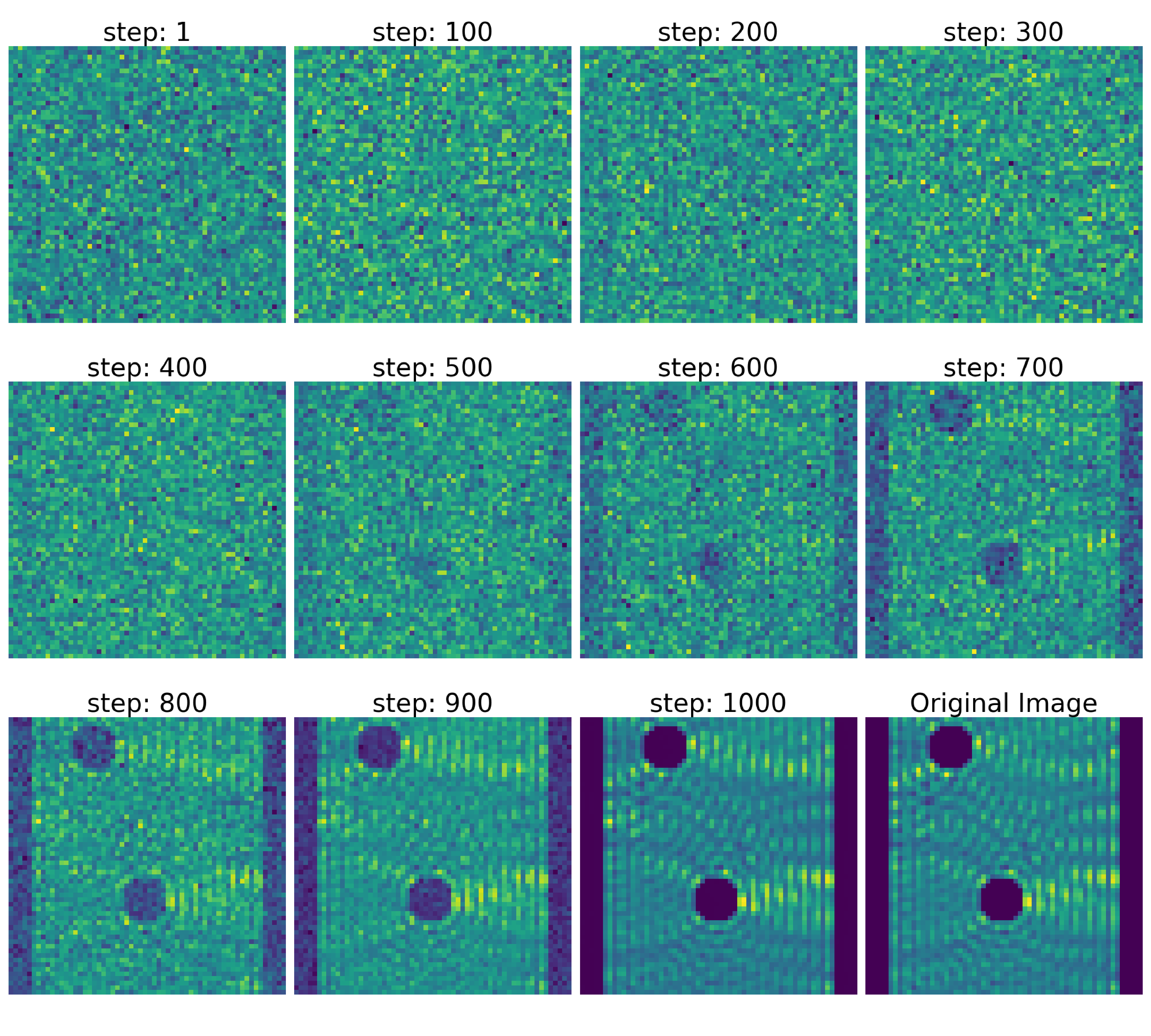}
\caption{Process for generating quantum interference pattern through 1000 generation steps.}  
\end{figure} 
The key mechanism by which this learning process is achieved is the self-attention 
mechanism discussed below.

\section{Self-attention as quantum interference extractor}   
The essence of the intricate fluctuations in magnetoconductance 
lies in the quantum interferences of wave functions scattered by defects and edges of the samples.   
For quantum interferences, the correlations of wave functions 
at two non-local points are essential.  
In our diffusion model, the self-attention mechanism built into the transformer blocks (red blocks in FIG. 2) plays a crucial role in incorporating these non-local correlations of wave functions. 

In the self-attention mechanism of machine learning, 
the attention weights that focus on relevant parts of the input data are calculated as follows \cite{vaswani2017attention}. 
First, three sets of vectors are created from the input data $\bm{x}$: the query  $\bm{q}$, 
the key $\bm{k}$, and the value $\bm{v}$. Then, the attention weight is calculated  
using the inner product of the vectors along the so-called channel direction, 
\begin{eqnarray}
A_{i j} = \frac{\exp \left[\bm{q}_{i}\cdot\bm{k}_{j}/\sqrt{d}\right]}{\sum_{j=1}^{d} \exp 
\left[\bm{q}_{i}\cdot\bm{k}_{j}/\sqrt{d}\right]}\, ,
\end{eqnarray}
where $d$ is the dimension of the vectors $\bm{k}_{i}$ (and $\bm{q}_{i}$).   
Finally, the input data is transformed into a weighted sum using the attention weights, 
\begin{eqnarray}
\tilde{x}_{i} = \sum_{j=1}^{d} A_{i j}\, v_{j} .
\label{eq: attention sum}
\end{eqnarray} 

The input data in our diffusion model consists of images of the quantum interference patterns: 
$x_{i} \propto \bigl| \psi(r_{i}) \bigr|^{2}$, where $r_{i}$ represents the position (or pixel) 
in the images. In this case, the attention weight $A_{i j}$ incorporates 
the correlation of wave functions at two distant points, allowing the self-attention mechanism 
to efficiently extract features of the quantum interferences.   
Moreover, by concatenating the quantum fingerprint data with the input data, 
it is possible to learn the information regarding non-local correlations 
together with the magnetoconductance,  
enabling the accurate generation of the quantum interference patterns.    

Through the learning process in the diffusion model, 
the attention weights for the interference images are obtained at each layer of the U-Net. 
We focus on the attention weights at the most downsampled layer (the third layer in FIG. 2), 
which contains relevant information regarding the importance of 
each pixel in the model.\footnote{For other layers, we implemented the Linear Attention \cite{katharopoulos2020transformers} to reduce the computational load.}  
From the attention transformation (\ref{eq: attention sum}), $B_{j} = \sum_{i=1}^{D} A_{i j}$  
corresponds to the relative importance of the pixel in the images before the transformation  
($\bm{v} \simeq \bm{x}$). 
The visualization of this importance, $B_{j}$, is depicted in FIG. 4. 
\begin{figure}[ht]
\includegraphics[width=8.6cm]{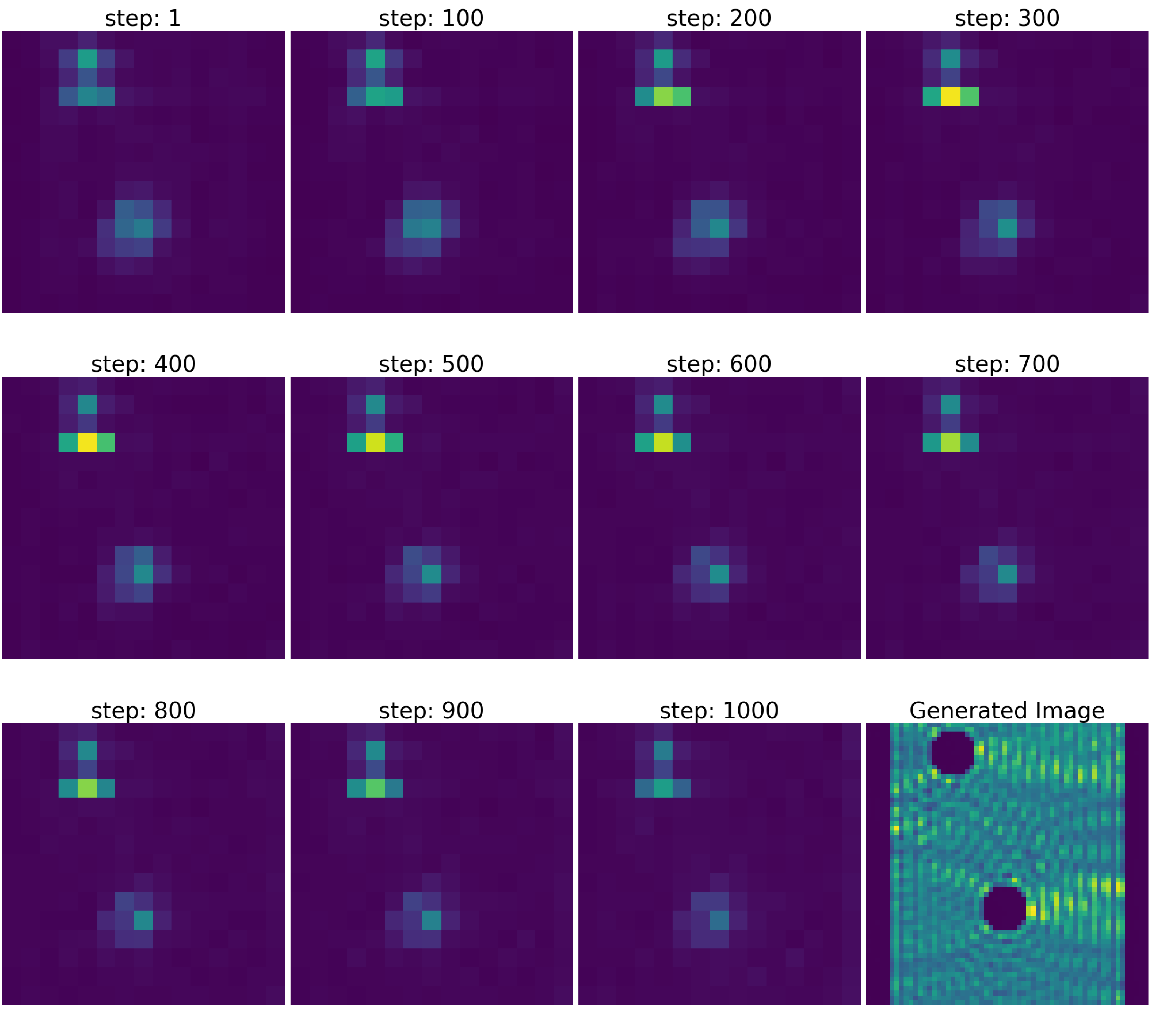}
\caption{Visualization of the attention weights in the first head through the generation process.}
\end{figure}
Compared to the generation process of the wave-function image shown in FIG. 3, 
we found that the importance of the defect locations and their vicinity 
has large values even at the stage when defects have not yet appeared in the wave-function image.
This demonstrates that the diffusion model can efficiently learn the long-distance 
correlations in the wave-function images through the self-attention mechanism 
and predict the positions of defects based on the conditioning from the magnetoconductance. 

Note that the self-attention mechanism in the diffusion model is multi-head attention 
with four heads, and each head exhibits similar importance for the antidot positions 
in a complementary manner.  
The first and second heads have relatively large values of importance 
for the upper antidot (FIG. 4), 
the third head has the similar values for both antidots, and  
the fourth head has a larger value for the lower antidot.

\section{Force fields in wave-function space}   
Through the denoising process of the diffusion model, 
the score functions in the data space,  
$s(\bm{x}) = \nabla_{\bm{x}} \log P(\bm{x})$, are obtained as discussed above.   
In our analysis, the learning data consists of wave-function images, $x_{i} \sim \bigl| \psi(r_{i}) \bigr|^{2}$, 
and thus the data space represents the function space of the wave functions: 
$\bm{{\cal X}} = \left\{ \bigl|\psi(\bm{r})\bigr|^{2} \right\}$.  
Note that the prepared wave-function images used as the learning data correspond to 
the genuine (numerical) solutions 
of the Schr\"{o}dinger equation for the corresponding antidot configurations.   

As an example, FIG. 5 illustrates the score function of the probability distribution 
in wave-function space 
around the wave function corresponding to the original image shown in FIG. 3. 
Since the wave-function space is a 3600-dimensional space 
composed of the wave-function images, 
the score function is visualized on a two-dimensional plane by projecting it onto two selected pixels 
for four different choices.  
Utilizing the trained diffusion model, the score function is obtained by 
estimating the noise from 50 noisy data points, each with 100 steps of added noise. 
\begin{figure}[ht]
\includegraphics[width=8.6cm]{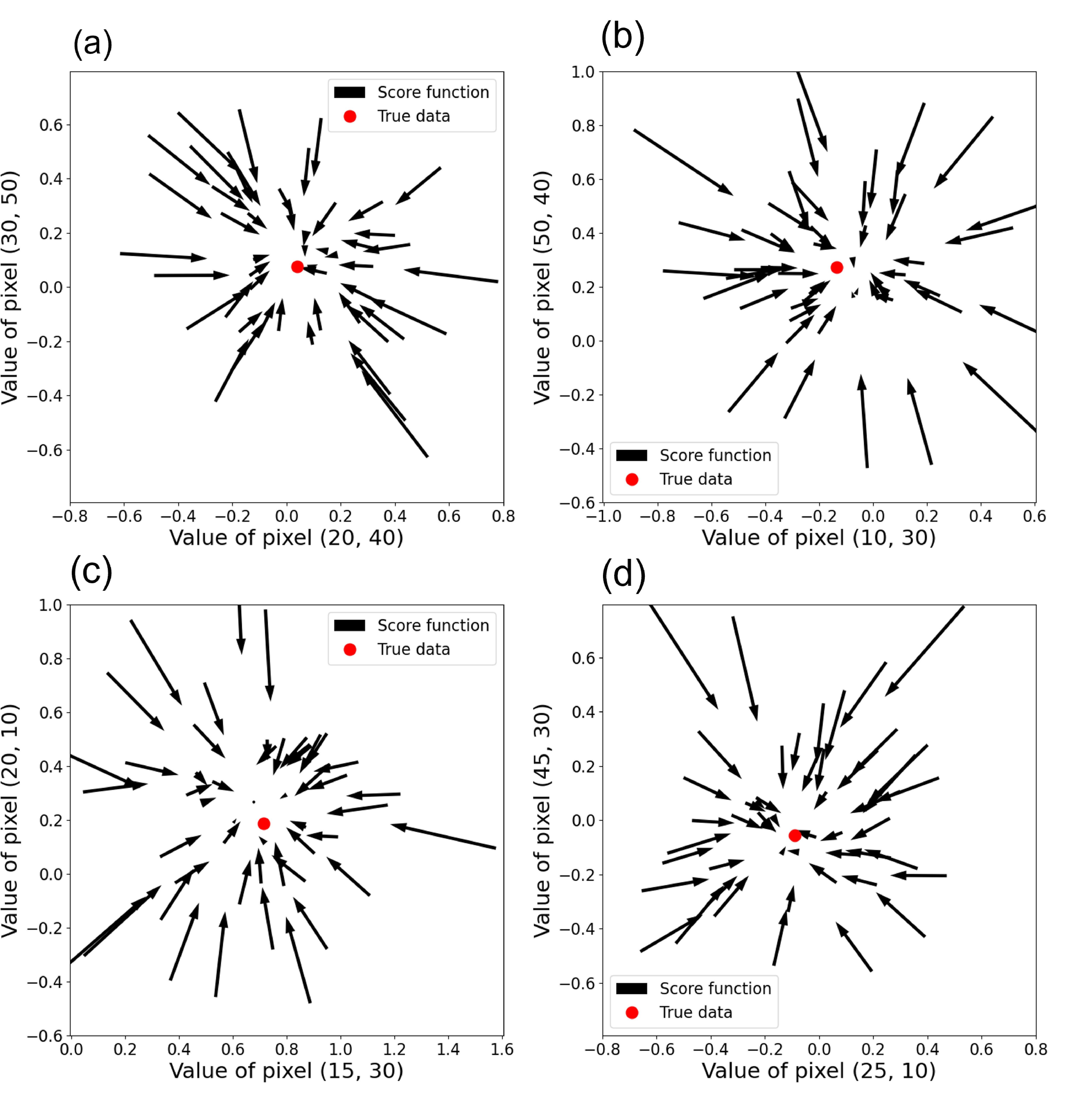}
\caption{Visualization of score function in different two-dimensional subspaces 
specified by two pixels: 
(a)  (20, 40) and (30, 50), (b) (10, 30) and (50, 40), (c) (15, 30) and (20, 10), (d) (25, 10) and (45, 30).}
\end{figure}

As seen in FIG.\,5, the score function, which represents the gradients of 
the probability distribution as two-dimensional vectors, 
is oriented towards the original (true) data from all noisy data points.  
Even when focusing on different pairs of pixels, the score function consistently points 
towards the original data. 
We have confirmed this property of the score function, which is generally expected in 
diffusion models, for the wave-function image data using this visualization.   
Furthermore, the energy functional in the wave-function space can be obtained from the score function 
via $s(\bm{x}) = - \beta\,\nabla_{\bm{x}} E(\bm{x})$. This visualization indicates that the energy 
functional has a (local) minimum at the true wave-function data and increases continuously 
as noise perturbations are added. Since the original image is the solution 
to the unperturbed Schr\"{o}dinger equation, this suggests that the denoising process of 
the diffusion model corresponds to linear response theory in quantum statistical physics, 
and provides a consistent energy functional in the wave-function space, at least locally, 
in the context of second quantization (or quantum field theory) \cite{altland2010condensed}.

\section{Summary}  
To understand the nature of quantum fingerprints in quantum transport, 
we have constructed the conditional diffusion model for machine learning of quantum 
interference patterns under the conditions of magnetoconductance. 
The diffusion model can generate the wave-function images corresponding to 
the magnetoconductance data with high accuracy, and the resulting 
attention weights, which efficiently extract information about non-local correlations 
of the wave-function, are visualized.  
Furthermore, the score functions in the wave-function space have been  
visualized, confirming the expected properties from the perspectives of machine learning 
and quantum statistical physics.    
In this letter, for simplicity, we have focused on the local structure around 
a single wave-function data point. However, the diffusion model is expected to retain 
more global information of the wave-function space, 
and the further research will provide more detailed insights into the structure of 
the wave-function space that emerges in quantum transport, from the perspective 
of quantum many-body systems (or quantum field theories). 

\begin{acknowledgements}
The authors thank Y. Hirasaki and K. Inui for useful discussions. 
This work was supported by
Institute of AI and Beyond of the University of Tokyo, Japan.
This work was also partially supported by 
CREST (No. JPMJCR20C1, No. JPMJCR20T2) from JST, Japan; 
Grant-in-Aid for Scientific Research (S) (No. JP19H05600), 
Grant-in-Aid for Transformative Research Areas (No. JP22H05114), and 
Grant-in-Aid for Early-Career Scientists (No. JP24K16985) from
JSPS KAKENHI, Japan.
\end{acknowledgements}  

\bibliographystyle{apsrev4-2}

\end{document}